\documentclass[prb,superscriptaddress,shownopacs,twocolumn,papersize=a4paper]{revtex4}
\usepackage{amsmath}
\usepackage{amsfonts}
\usepackage{amssymb}
\usepackage{graphicx}
\usepackage[usenames]{color}
\usepackage[normalem]{ulem}

\begin{document}

\title{Spectral non-uniform temperature, non-local heat transfer, and the
spin Seebeck effect}
\author{Konstantin S. Tikhonov}
\affiliation{Department of Physics, Texas A\&M University, College Station, TX
77843-4242, USA}
\email{tikhonov@physics.tamu.edu}
\affiliation{Department of Condensed Matter Physics, The Weizmann Institute of Science,
Rehovot 76100, Israel}
\affiliation{L. D. Landau Institute for Theoretical Physics, 117940 Moscow, Russia}
\author{Jairo Sinova}
\affiliation{Department of Physics, Texas A\&M University, College Station, TX
77843-4242, USA}
\affiliation{Institute of Physics ASCR, v.v.i., Cukrovarnick\'a 10, 162 53 Praha 6, Czech
Republic}
\author{Alexander M. Finkel'stein}
\affiliation{Department of Physics, Texas A\&M University, College Station, TX
77843-4242, USA}
\affiliation{Department of Condensed Matter Physics, The Weizmann Institute of Science,
Rehovot 76100, Israel }

\begin{abstract}
We present a theory of the spin Seebeck effect driven by subthermal
non-local phonon heat transfer and spectral non-uniform temperature
distribution.  The theory explains the non-local behavior of the effect
arising from the fact that phonons that store the
energy (thermal) and the phonons  that transfer it (subthermal) are located
in different parts of the spectrum and have very different kinetics.  This
gives rise to a spectral phonon distribution function that deviates from
local equilibrium along the substrate and is sensitive to boundary conditions.
The theory also predicts a
non-magnon origin of the effect in ferromagnetic metals in agreement with observations in recent experiments.
Equilibration of the heat flow out of
the substrate to the Pt probe and backwards leads  to a measurable vertical
spin-current produced by the spin polarized electrons dragged by the local
thermal phonons. We predict specific sample length limits and other
dependencies that can be probed experimentally, and obtain the correct
magnitude of the effect.
\end{abstract}

\maketitle

\begin{figure}[h]
\includegraphics[width=0.95\columnwidth,angle=0]{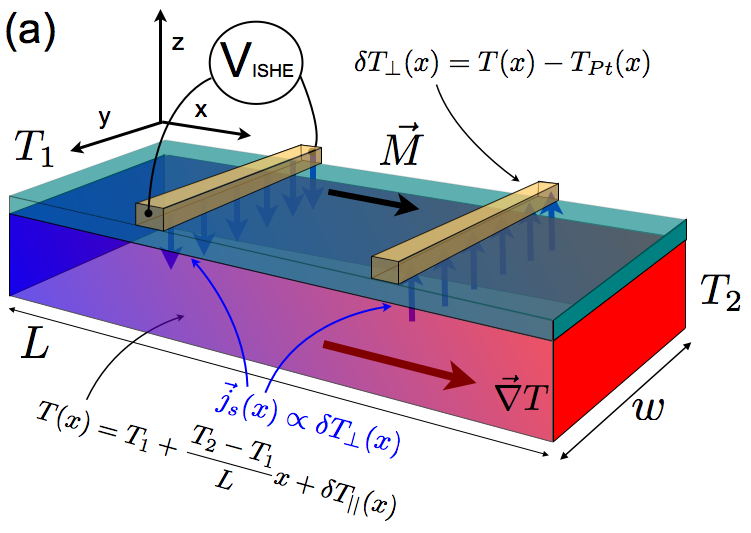} %
\includegraphics[width=0.95\columnwidth,angle=0]{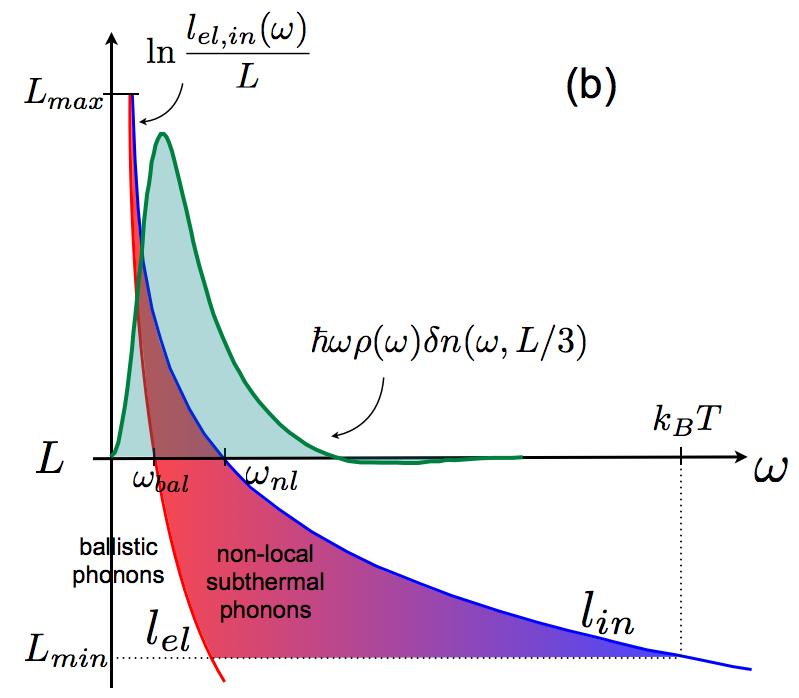}
\caption{(a) Scheme of the SSE experiment. The effect incorporates three key
physical mechanisms: (i) subthermal phonons whose inelastic length, $l_{in}$%
, is of the order of the sample size, $L$, and whose elastic length, $l_{el}$%
, is smaller than $L$, drive the non-local heat propagation along the
substrate which gives rise to a steady state distribution function that
deviates from local equilibrium; (ii) equilibration of heat flows out of the
substrate into the Pt probe and backwards establishes {the temperature in
the probe $T_{Pt}\ne T(x)$}; (iii) the different phonon distribution
functions in the probe and the substrate yield a \emph{spin}-phonon-drag
current, $\vec{j}_s(x)\propto\protect\delta T_{\perp }$. (b) Spectral phase
diagram of phonons as a function of sample length. The deviation from local
thermal equilibrium, $\protect\delta n(\protect\omega,x)=n(\protect\omega%
,x)-n_{T(x)}(\protect\omega)$, is illustrated (green curve) for $x=L/3$;
here $\protect\rho_{ph}(\protect\omega)\propto \protect\omega^2$ is the
phonon density of states.}
\label{Fig1}
\end{figure}
Recently discovered spin-dependent thermoelectric effects have opened the
doors to a new frontier of spintronics that merge spin, charge, and thermal
physics, known as \textit{spin caloritronics}.\cite%
{Uchida08,jaworski10,jaworski11,Uchida10,adachi10,uchida11,Flipse2012,Costache2012,Bauer2012}
The key and most puzzling effect among them is the transverse spin Seebeck effect (SSE)
in which a thermal gradient in a ferromagnet/substrate structure gives rise
to spin-currents which vary along the length of the sample and are detected
via the inverse spin-Hall voltage.\cite{valenzuela06} This effect has been
experimentally observed using different ferromagnetic materials: metals,\cite%
{Uchida08} semiconductors,\cite{jaworski10,jaworski11} and insulators.\cite%
{Uchida10} The magnitude of the SSE is quantified by the transport
coefficient $S_{xy}=\frac{V_{y}}{w\nabla _{x}T}$, where $V_y$ is the
measured ISHE voltage, $w$ is the width of Pt probe, and $\nabla
_{x}T=(T_2-T_1)/L$, where $L$ is the length of the sample, see Fig.~\ref%
{Fig1}(a). The ISHE voltage is given by ${V_y}=\frac{2\left\vert
e\right\vert \rho \theta _{H}}{\hbar }(\mathbf{j}_{s}\mathbf{\times s})_y,$
where $\mathbf{j}_{s}$ is the spin current, $\mathbf{s}$ is its
polarization, $\theta _{H}$ is the spin-Hall angle of the probe (in Pt, $%
\theta _{H}$ is of the order of one per cent) and $\rho $ is its electric
resistivity. The effect is non-local, \textit{i.e.} it depends on the
position along the sample rather than the local temperature gradient. In
addition, the size of the sample is usually about $1$ $cm$ and such a
long-ranged information about position can be transferred only by phonons,
propagating along the insulating substrate.\cite{jaworski10}

Here we show that the non-local SSE is
a consequence of the non-local energy transfer due to sub-thermal diffusive phonons that are sensitive to the boundary conditions and give rise to a spectral non-uniform temperature along the sample.\cite{levinson80, levinson80b}.
In addition, we
demonstrate that while in the insulator the SSE is likely determined by the
phonon-magnon mechanism, in the conducting ferromagnet (e.g., Ni$_{81}$Fe$%
_{19}$\cite{Uchida08} and GaMnAs\cite{jaworski11}), the magnon mechanism is
not the only one available. In fact, in recent measurements in bilayer F-Pt
wire devices,  the specific geometry excludes long-ranged propagation of
magnons  and leaves only phonons as a source of non-locality.\cite{uchida11}
In addition, the experiments by Jaworski \textit{et al.} in Ref.~%
\onlinecite{jaworski11} were performed on a material with Curie temperature $%
T_{C}=130$ K, considerably lower than the Debye temperature $\theta _{D}=350$
K, and showed that $V_{ISHE}\propto M$ at the vicinity of the Curie point;
i.e. the SSE signal vanishes with the magnetization $M$ {with} the same
critical behavior. This latter fact excludes the magnon mechanism for this
case.

The theory of this phonon-electron SSE, which does not involve magnons,
has three key physical mechanisms. The first (i) involves
the non-local nature of the signal driven by subthermal phonons. %
{In recent measurements of the SSE in
insulators\cite{agrawal12} the temperature difference between thermal
magnons  and phonons assumed in the current theory\cite{xiao10} has not been
observed, suggesting the necessity of the concept of spectrally non-uniform
temperature.} This concept originates from the fact that in most
dielectrics, and also some semiconductors, the energy transfer is highly
non-local\cite{levinson80, levinson80b} because of the strong dependence of
the diffusion coefficient of phonons on frequency: $D\left( \omega \right)
\propto \omega ^{-4}$, if the dominant scatterers are point-like \cite%
{ziman01}. In the diffusive regime of the experiments the energy relaxation
length is given by $l_{in}\left( \omega \right) =\sqrt{D\left( \omega
\right) \tau _{in}\left( \omega \right) },$ where the energy relaxation rate
is $\tau _{in}^{-1}\left( T,\omega \right) \propto T^{4}\omega .$ While the
thermal phonons, $\hbar \omega \sim k_B T$, are equilibrated, the subthermal
low-frequency phonons can deviate from the local equilibrium due to the
rapid low-frequency growth of inelastic length $l_{in}\left( \omega \right)
=l_{in}\left( T\right) \left( T/\omega \right) ^{5/2},$ which leads to
non-local kinetics. Even the concept of the temperature itself is
well-defined only for phonons of high enough frequency. For the 'thermal'
part of the spectrum $\hbar \omega \gtrsim k_B T$, the distribution function
has a Planckian form $n_{T}\left( \omega\right) =\left( e^{\hbar \omega /
k_B T}-1\right) ^{-1}$ with a local temperature $T=T\left( x\right) $. As a
result, \emph{the phonons which store the energy and phonons which transfer
it are located in different parts of the spectrum. } {As illustrated in Fig.~%
\ref{Fig1}(b),} this spectral separation occurs when $l_{el}\left( T\right)
\ll l_{in}\left( T\right) \ll L,$ where $l_{el}\left( T\right) \equiv
l_{el}\left( \hbar \omega =k _B T\right) \propto D(\hbar \omega=k_B
T)\propto T^{-4},$ and $l_{in}\left( T\right) \equiv l_{in}\left(\hbar
\omega =k_B T\right) \propto T^{-4.5}$. Then the subthermal phonons whose
inelastic length is of the order of {(but elastic length is much shorter)
than} the sample size $L$ drive the non-local heat propagation along the
substrate, giving rise to a steady state phonon distribution function that
deviates from local equilibrium for $\hbar \omega\ll k_B T$ and depends on
the position along the substrate.  {To describe this non-local effect, it is
essential to formulate the boundary conditions for the equations describing
the propagation of the diffusive phonons, see below.}

The second (ii) mechanism involves the {electron-phonon drag}. Since the
probe is a 'dead end', there is a full balance between incoming and outgoing
heat fluxes such that net heat flux is zero. However, the incoming and
outgoing fluxes have different spectral distributions, because of the
inelastic processes in Pt which average out the spectrum of the incoming
flux, and establish a local temperature ${T}_{Pt}(x)$ different from $%
T\left( x\right) $. The spin drag, induced by the phonon flux, is sensitive
to the spectral content of the phonon distribution function. Hence, despite
zero net heat flux, the spin injection is not zero. In the stationary
situation, the drag voltage induced by the phonons is compensated by
redistribution of the electron density, so that the total electric current
is zero (as well as electrochemical potential gradient). However, in the
presence of a spin polarization, there will be a net spin current $%
j_{s}=j_{\uparrow }-j_{\downarrow }$ polarized along magnetization $M$:
unlike its charge counterpart, {spin drag} is not blocked by accumulation of
the spin density, {which is eliminated by SO interaction in Pt}. The
magnitude of $j_{s}$ depends on the ratio of the thickness of the
ferromagnet, $d_{F},$ and the phonon inelastic scattering length there, $%
l_{in}^{F}$. The optimal value of $d_{F}$ for observing the {phonon drag}
SSE is of the order of $l_{in}^{F}\left( T\right)$. For too thin
ferromagnet, $d_{F}\ll l_{in}^{F}$, the phonons cannot effectively transfer
their momentum to electrons to drag them toward the probe. In the opposite
limit, $d_{F}\gg l_{in}^{F}$, the phonons equilibrate before they reach the
region near the probe. {(An alternative mechanism not considered here is the quantum
acoustoelectric pumping\cite{levinson00} due to the spectrally non-uniform flux of phonons.\cite{future})}

The final (iii) mechanism involves the conversion of the spin-current to an
electric signal via the ISHE. This conversion is most optimal if the
thickness of the Pt layer is of the same order of magnitude as the spin
relaxation length in Pt, which is the case in the discussed experiments.\cite%
{zhang12} As shown in detail in the following sections, the resulting theory
gives the correct magnitude of the signal, predicts a dependence on
magnetization $S_{SSE}\propto M$, and gives specific temperature and size
dependencies that can be tested experimentally.

\section{\textbf{Results}}

\subsection{Subthermal phonon kinetics}

On Fig. \ref{Fig1}(b) we show the spectral phase diagram of frequency
regions contributing differently to the kinetics of phonons. There are two
characteristic frequencies, $\omega _{nl}$ and $\omega _{bal}$, determining
the propagation of phonons:
\begin{equation}
l_{in}\left( \omega _{nl}\right) =L,~~~l_{el}\left( \omega _{bal}\right) =L.
\label{ws}
\end{equation}
For non-local transport we require $\hbar \omega _{nl}=k_B T\left(
L/l_{in}\left( T\right) \right) ^{-2/5}\ll k_B T$. In addition, we will not
be interested in phonons in the ballistic part of the spectrum, $\omega
<\omega _{bal}$. This is legitimate as long as $\omega _{bal}\ll \omega
_{nl},$ and determines a maximum length of the sample, $L_{\max }$, given by
the point of intersection of the curves $l_{in}\left( \omega \right) $ and $%
l_{el}\left( \omega \right) $, as shown Fig.~\ref{Fig1}(b). This gives a
temperature dependence $L_{\max}\propto T^{-16/3}.$ For lengths larger than $%
L_{\max}$, the non-local effect is due {to the fraction of
phonons propagating ballistically} and requires a different formalism, which
we will not discuss here. The other condition that allows to separate
thermal phonons from those which produce non-local effects is $l_{in}\left(
T\right) \ll L$. This gives a minimum length of the sample $L_{\min }\propto
T^{-9/2}$. {For length smaller than $L_{min}$ even thermal phonons are out
of equilibrium and spectral separation does not hold.}  The large ratio of $%
l_{in}\left( T\right) /l_{el}\left( T\right) $ opens the window $L_{\min
}\ll L\ll L_{\max },$ which we are interested in. Hence the sample size
should be in the range indicated on Fig.~\ref{Fig1}(b). Estimation at $T=10$
K (when typical phonon energy is $28$ K), gives $L_{\max }$ about few $cm$
and $L_{\min}\ $ on the scale of $mm$. Recall that the typical size of the
sample used for the SSE experiments is $1cm$. {With
temperature the width of the region of applicability of the theory behaves
as $L_{max}/L_{min}\propto T^{-5/6}$ and we expect it to be relevant up to 50K.
In addition}, the temperature is assumed to be much smaller than the Debye
temperature, $T\ll \theta _{D}$, which allows us to ignore Umklapp processes.

With these specific length restrictions we consider next the theory of
propagation of diffusive phonons along the substrate. Owing to the fact that
the low-frequency phonons do not primarily interact with themselves but with
equilibrated high-frequency phonons, one may use the following kinetic
equation that describes propagation of phonons in the insulating substrate,
valid for $\hbar \omega \lesssim k_B T$:
\begin{equation}
D\left( \omega \right) \partial _{x}^{2}n\left( \omega ,x\right) =\frac{%
\delta n\left( \omega ,x\right) }{\tau _{in}\left( \omega \right) },
\label{ke}
\end{equation}
where $\delta n$ is the deviation from the local equilibrium
\begin{equation}
\delta n\left( \omega ,x\right) =n\left( \omega ,x\right) -n_{T(x)}(\omega).
\label{dloc}
\end{equation}
To get a closed set of equations, one needs an equation for $T\left(
x\right) $, which deviates from the linear behavior due to the non-locality
of the heat transport. This equation is obtained from the continuity of the
energy density in the system, which in stationary situations reads as $%
\nabla\cdot \vec{j}_Q=0$.

Because of the divergence of $D\left( \omega \right) $ at small $\omega ,$
the heat flux $\vec{j}_Q$ is transported by the low-energy part of the
spectrum.\cite{peierls29, pomeranchuk41, herring56} The heat current density
is given by:
\begin{equation}
\vec{j}_Q\left( x\right) =-\int_{0}^{\infty }\hbar \omega \rho_{ph} \left(
\omega \right) D\left( \omega \right) \mathbf{\partial }_{x}n\left( \omega
,x\right) d\omega ,  \label{qr}
\end{equation}
where $\rho_{ph} \left( \omega \right) \propto \omega ^{2}$ is the phonon
density of states (summed over all branches). The integral for $\vec{j}_Q(x)$
diverges and has to be cut off at small frequency (the exact value of the
cut off does not enter our results since the integral for $\nabla\cdot \vec{j%
}_Q$ converges). Using Eq. (\ref{ke}), the energy density continuity
equation takes the following form:

\begin{equation}
\int_{0}^{\infty }\omega ^{2}\delta n\left( \omega ,x\right) \rho_{ph}
\left( \omega \right) d\omega =0.  \label{Tx}
\end{equation}
This equation should hold for all $x$. Thus, one has to solve a system of
integro-differential equations. For the case of a pulse propagation in an
infinite media the non-local phonon transport has been studied in Ref. %
\onlinecite{levinson80, levinson80b, wilson84}. However, we are interested
in a stationary solution in the presence of the boundaries.

On the boundary between the substrate and the heater there is a jump in the
phonon distribution function, because of the abrupt change in the properties
of materials. This leads to a finite thermal boundary resistance (Kapitza
resistance), which manifests itself through the jump $\Delta T_{K}$ at the
contact.\cite{swartz89} If the scattering in the vicinity of the boundary is
mostly elastic, the boundary condition consists of conservation of spectral
heat current density across the boundary. {It relates the heat flux through
the boundary to the jump of the phonon distribution function across it.} At
the left end of the sample (which is at heat contact with a reservoir at
temperature $T_{1}$) {it takes the following form}:
\begin{equation}
l_{el}\left( \omega \right) \left. \partial _{x}n\left( \omega ,x\right)
\right\vert _{x=0}=\frac{1}{R_{Bd}}\left[ n\left( \omega ,0\right)
-n_{T_{1}}\left( \omega \right) \right] .  \label{bc}
\end{equation}
The boundary resistance $R_{Bd}$ is assumed to be frequency independent. {If the heat contacts are in thermal equilibrium, $R_{Bd}$} can be related to the thermal boundary conductivity $h_{Bd}=\frac{\dot{Q}}{%
A\Delta T_{K}}\propto \frac{T^{3}}{v_s^{2}}R_{Bd}^{-1},$ where $v_s$ is the
averaged sound velocity. {Note, that in the absence of the boundary resistance ($R_{Bd}=0$), the locally equilibrium distribution function $n(\omega,x)=n_{T_0(x)}(\omega)$ with $T_0(x)=T_1+(T_2-T_1)x/L$ satisfies both the kinetic equation and the boundary conditions, so that $\delta n=0$ and the non-local effect vanishes. One may easily see that $\delta n\propto R_{Bd}$ at not too large values of $R_{Bd}$.}
\begin{figure}[tbp]
\includegraphics[width=220pt,angle=0]{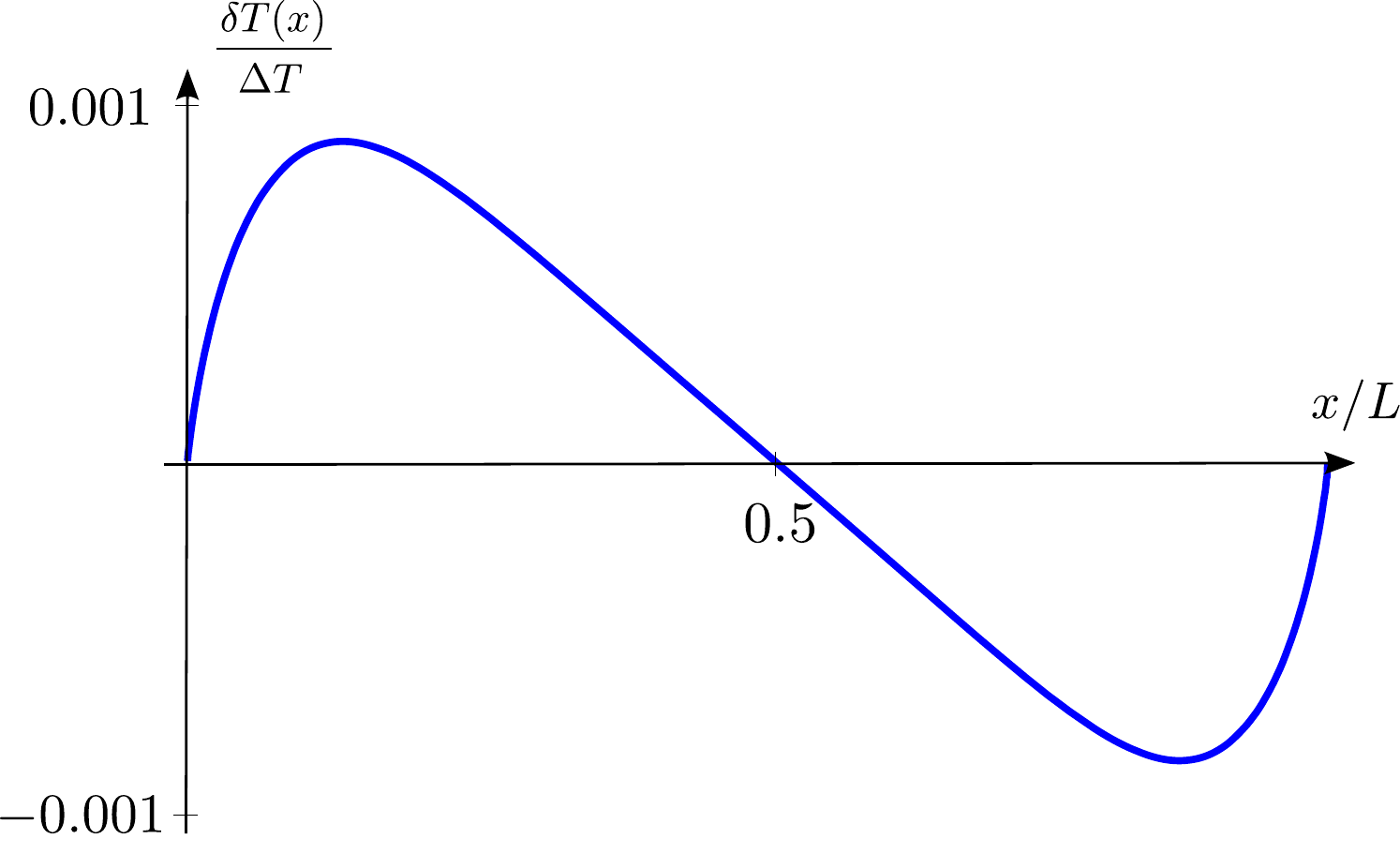}
\caption{Correction to the linear temperature dependence as a function of
position, Eq.~(\protect\ref{linearT}).}
\label{fig_temp}
\end{figure}

With this it is then possible to write down a closed equation for $T\left(
x\right) $. If the phonon temperature as a function of position $x$ is
known, the distribution function {can be obtained from Eq. (\ref{ke}) and}
reads:
\begin{gather}
n\left( \omega ,x\right) =n^{\left( S\right) }\left( \omega ,x\right) +
\label{Bx} \\
+Z_{\omega }\int \Pi _{\omega }\left( L-x_{<}\right) \Pi _{\omega }\left(
x_{>}\right) n_{T\left(x^{\prime }\right) }\left( \omega \right) dx^{\prime
}/l_{in}\left( \omega \right) ,  \notag
\end{gather}
where $x_{<}=\min \left( x,x^{\prime }\right) ,~x_{>}=\max \left(
x,x^{\prime }\right) .$ The 'source' term $n^{\left( S\right) }\left( \omega
,x\right) $ comes from the boundary condition (\ref{bc}) and is equal to
\begin{equation}
n^{\left( S\right) }\left( \omega ,x\right) =g_{\omega }Z_{\omega }\left[
n_{T_{1}}\left( \omega \right) \Pi _{\omega }\left( x\right)
+n_{T_{2}}\left( \omega \right) \Pi _{\omega }\left( L-x\right) \right] ,
\label{ns}
\end{equation}
where
\begin{equation}
\Pi _{\omega }\left( x\right) =\cosh \left( \frac{L-x}{l_{in}\left( \omega
\right) }\right) +g_{\omega }\sinh \left( \frac{L-x}{l_{in}\left( \omega
\right) }\right) ,  \label{pi}
\end{equation}
and
\begin{equation}
Z_{\omega }=\left[ 2g_{\omega }\cosh \left( L/l_{in}\left( \omega \right)
\right) +\left( 1+g_{\omega }^{2}\right) \sinh \left( L/l_{in}\left( \omega
\right) \right) \right] ^{-1}.  \label{z}
\end{equation}
Above we have introduced the effective boundary thermal conductance $\
g_{\omega }=\left( l_{in}\left( \omega \right) /l_{el}\left( \omega \right)
\right) R_{Bd}^{-1}$. The second term in Eq. (\ref{Bx}) describes the
process of redistribution of phonons along the sample due to diffusion and
inelastic scattering. Substituting Eq. (\ref{Bx}) into Eq. (\ref{Tx}), one
gets an integral equation for $T\left( x\right) $, which can be solved
numerically. This procedure is self-consistent: after finding $T\left(
x\right) $, the distribution function is easily calculated from Eq. (\ref{Bx}%
). To illustrate the result, we assume the following ratios of
characteristic lengths of a thermal phonon
\begin{equation}
L:l_{in}\left( T\right) :l_{el}\left( T\right) =1:0.1:0.005,  \label{rel}
\end{equation}
and calculate the correction $\delta T_{\parallel }\left( x\right) $ to the
linear temperature behavior
\begin{equation}
T\left( x\right) =T_{1}+\frac{T_{2}-T_{1}}{L}x+\delta T_{\parallel }\left(
x\right) ,  \label{linearT}
\end{equation}
which is shown on Fig. \ref{fig_temp}, where we have assumed that $R_{Bd}=0.1
$ {and $T_1 < T_2$}. Although the deviation from the linear behaviour is
small, it
{ensures the conservation of the energy density of the phonons propagating along the substrate.
Ultimately, the non-equilibrium correction $\delta n\left( \omega ,x\right)$ is responsible for the SSE effect.}
On Fig.~\ref{Fig1}(b), the frequency dependence of $\hbar \omega \rho_{ph}
\left( \omega \right) \delta n\left( \omega ,x\right)$ is plotted close to
the colder end (for $x=0.3L)$. On the hotter end,~$\delta n$ has the
opposite sign.

\subsection{Out of plane spin transport}

After finding the non-equilibrium distribution function of phonons $\delta
n\left( \omega,x\right) $, we next concentrate on the heat and spin
transport in the vertical direction from the substrate to the probe across
the magnet. The temperature of phonons in the Pt probe, ${T}_{Pt}\left(
x\right)$, is different from the $T\left( x\right)$. It is determined by the
requirement that the heat flux created by non-equilibrium non-local phonons
from the substrate to Pt is compensated by back-flow flux of thermal phonons
from Pt to the substrate. The resulting temperature difference, $\delta
T_{\perp }\left( x\right) =T\left( x\right) -{T}_{Pt}\left( x\right)$, can
be found from the heat balance equation:
\begin{equation}
\int_{0}^{\infty }\omega \rho_{ph} \left( \omega \right) \delta N\left(
\omega ,x\right) d\omega =0,  \label{hb}
\end{equation}
where $\delta N\left( \omega ,x\right) =n_{sub}-n_{Pt}$ is the difference of
the distribution function of phonons entering and leaving the Pt probe,
located at $x$. Here we neglect inelastic scattering of phonons while they
pass through the ferromagnetic layer ($d_{F}\lesssim l_{in}^{F}$), and have
assumed the sound velocities to be of similar order in the Pt and the
substrate. We also assume, that the probe is small enough $a\ll l_{el}(T)$,
so that the influence of the counterflow on the phonon distribution function
in the substrate can be ignored.

\begin{figure}[tbp]
\includegraphics[width=220pt,angle=0]{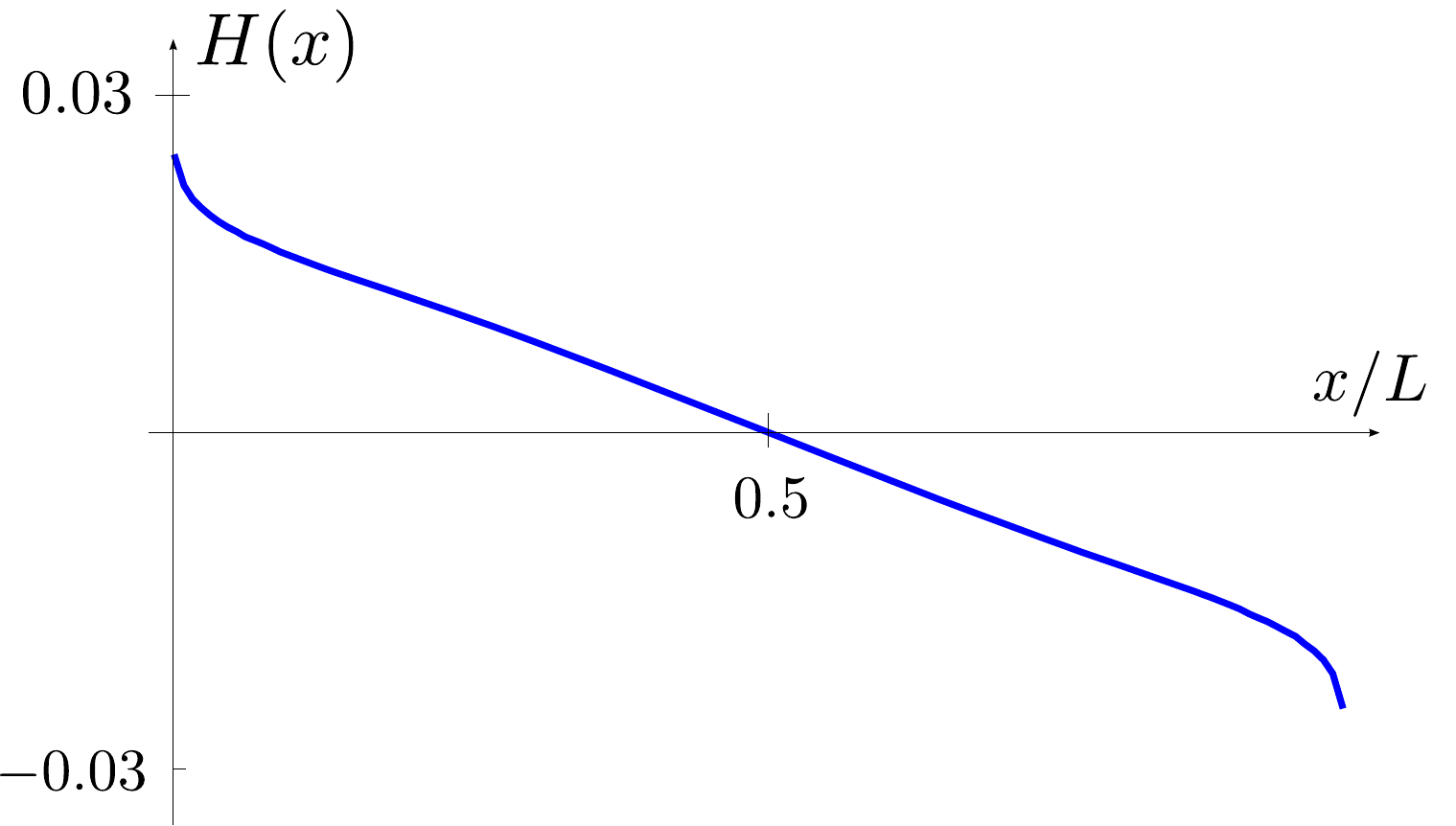}
\caption{The function $h(x)\propto \protect\delta T_{\perp }(x)$ determining
the magnitude and spatial profile of the SSE signal $S_{xy}$ given in Eq.~(%
\protect\ref{res}).}
\label{fig_dt}
\end{figure}

It is useful to present $\delta N\left( \omega ,x\right) $ in the
following form:
\begin{equation}
\delta N\left( \omega ,x\right) =\left[ n_{T\left( x\right) }\left( \omega
\right) -n_{T_{Pt}\left( x\right) }\left( \omega \right) \right] +\delta
n\left( \omega ,x\right) .  \label{dN}
\end{equation}%
Then, the the temperature difference $\delta T_{\perp }\left( x\right) $ can
be calculated from the equation:
\begin{equation}
\delta T_{\perp }\left( x\right) /T\left( x\right) \propto -\int_{0}^{\infty
}z^{3}\delta n\left( zT,x\right) dz\equiv -h\left( x\right) ,  \label{bal}
\end{equation}%
where $h\left( x\right) $ is the dimensionless heat flux supplied to the
probe by the nonequilibrium phonons. {The function $h\left( x\right) $ can be
written in a form of $h\left( x\right) =$ $\frac{\Delta T}{T}\left( \omega
_{bal}/T\right) ^{4}H\left( x\right),$ where $H\left( x\right) $ is a slow
function of temperature and boundary resistance $R_{Bd}$. Here $\omega_{bal}$ encodes information about scattering of phonons on the disorder and
the length of the sample. Function $H(x)$ is plotted on Fig.~\ref{fig_dt} for the same sample
parameters as before.}

With this we can finally estimate the scale of the SSE due to conducting
electrons, dragged by out-of-equilibrium phonons, in more detail. The
guiding idea about the scale of the effect follows from the derivation of
the well known Gurevich formula\cite{gurevich46, blatt76, lifshitz81} for
the phonon drag. This formula gives for thermoelectric coefficient $\eta
=-j/\nabla T$ the following expression: $\eta \propto -\frac{\sigma T^{3}}{%
e\left( p_{F}v_{s}\right) ^{3}}$, which is valid when $q_{T}l\gg 1$ (here $%
q_{T}=k_{b}T/\hbar v_{s}$ is the wavelength of a thermal phonon). For the
dirty case $q_{T}l\ll 1,$ the particle current density dragged to the probe
is given by \cite{sergeev01}:
\begin{equation}
j_{e}^{z}\left( x\right) \propto \frac{\tau _{ei}}{p_{F}}\int_{0}^{\infty
}\omega ^{2}\delta N\left( \omega ,x\right) W\left( \omega l_{ei}/u\right)
\rho _{ph}\left( \omega \right) d\omega .  \label{pmp}
\end{equation}%
We write $\tau _{ei},~l_{ei}$ for electron-impurity scattering time and
length in the ferromagnet. The role of electron-impurity scattering in Eq. (%
\ref{pmp}) is twofold. It enhances electron-phonon interaction by slowing
the motion of electrons (making it diffusive). This is taken into account by
the form of $W(x)$. On the other hand it diminishes the drag effect due to
the loss of electron momentum by impurity scattering. The details of the
function $W\left( x\right) $ depend on the character of the scattering of
phonons on defects. We assume that phonons scatter on impurities vibrating
with the lattice\cite{pippard55} and $W\left( x\right) =W_{vb}\left(
x\right) $. For temperatures $T\ll u/l_{ei},$ we may use the asymptotic
behavior $W_{vb}\left( x\ll 1\right) \approx x$. Recalling that the charge
current will be compensated by an unpolarized backflow charge current from
the Pt probe, the total spin current is given by the polarized current
dragged by the phonons into the Pt probe. Finally, we rewrite the expression
for the spin-phonon-drag current injected into the Pt probe as
\begin{equation}
j_{s}^{z}\left( x\right) \propto X_{M}T\left( T/v_{s}\right) ^{2}\left(
T/\theta _{D}\right) ^{2}A_{el}\left( T\right) J\left( x,T\right) ,
\label{pmp2}
\end{equation}%
where $A_{el}\left( T\right) =\left( k_{B}T/\epsilon _{F}\right) \left(
k_{F}l_{ei}\right) ^{2}$ is a dimensionless constant, determined by
electrons, {$X_{M}=\frac{n_{\uparrow }-n_{\downarrow }}{n_{\uparrow }+n{%
\downarrow }}$ is the level of spin polarization} and the dragging factor is
\begin{equation}
J\left( x,T\right) =\int_{0}^{\infty }z^{5}\delta N\left( zT,x\right) dz.
\label{df}
\end{equation}%
Since the spectral densities of the energy and the charge currents are
proportional to different powers of the phonon frequency, the electronic
drag due to phonons is possible even when the net energy flow is zero. The
contribution {to J in Eq. (\ref{df})} arising due to the temperature
difference $\delta T_{\perp }\left( x\right) $ between the substrate and the
Pt probe (first term in Eq. (\ref{dN})) is dominant. In other words, while
the non-locality of the effect along the sample is carried by the
low-frequency phonons, the dragging force generating the spin-current is
produced by the thermal phonons. As a result of this intricate joint effort
by the phonons in different parts of the spectrum, one gets (restoring
units):
\begin{equation}
j_{s}^{z}\left( x\right) =X_{M}\left( k_{B}/\hbar \right) ^{3}A_{el}\left(
T\right) \left( T/\theta _{D}\right) ^{2}\left( T/v_{s}\right) ^{2}\delta
T_{\perp }\left( x\right) .  \label{pmpf}
\end{equation}%
Finally, for the magnitude of the SEE, $S_{xy}=-2\left( \left\vert
e\right\vert \rho /\hbar \right) \theta _{H}j_{s}^{z}/\nabla _{x}T$, and
recalling that $\delta T_{\perp }(x)\propto -Th(x)$, we obtain:
\begin{equation}
{S_{xy}=\theta _{H}S_{xy}^{\left( 0\right) }A_{el}\left( T\right)
X_{M}k_{F}l_{el}\left(\theta _{D}/2.8\right) H\left( x\right)} ,  \label{res}
\end{equation}
where $S_{xy}^{\left( 0\right) }=\left\vert e\right\vert k_{F}\rho $ is a
material-dependent constant. The factor $2.8$ takes into consideration that the
energy of thermal phonon is $2.8k_{B}T$. Assuming that in Pt, $\rho =0.9\,\,\mu \Omega
\cdot m~$and $k_{F}^{-1}=10^{-8}cm$, we obtain $S_{xy}^{\left( 0\right)
}\approx 30\mu $V/K. Function $H\left( x\right) $ is positive at the cold
end, meaning the dragging force pushes electrons towards the magnet there,
according to Eq. (\ref{pmpf}). {Note that (i) although the electron-phonon drag is proportional to a high power of temperature, see Eq. (\ref{pmp2}), the final result for the SSE coefficient is only weakly temperature dependent, $S_{xy}\propto T$. It comes out as a result of the strong dispersion of the phonon scattering time in the substrate. Although function $H(x)$ in Eq. (\ref{res}) is also temperature dependent, this dependence comes only from the non-locality of the phonon collision integral in energy, and is relatively weak. Another important property of this function is that (ii) it's spatial dependence varies with temperature rather slowly. This is because the phonons which contribute mostly to the non-local effect have inelastic scattering length of the order of the sample size. Varying the temperature mainly results in the shift of the relevant phonon energy $\omega_{nl}$, so that the corresponding length scale $l_{in}(\omega_{nl})$ remains the same. {\em The observations (i), (ii) stress the importance of the strong dispersion of the phonon scattering.}}

{Note that the sign of the Seebeck follows the sign of $H(x)$}
and is positive at the cold end. Taking $\theta _{H}=0.0037$, $%
\epsilon _{F}/k_{B}=10^{3}K,~\theta _{D}=350K,~k_{F}l_{ei}=10$ and the ratio
of characteristic lengths as in (\ref{rel}), we find the magnitude of the
effect at $10K$ to be $S\sim 1\mu $V/K$\times X_{M}$.

\section{Discussion and conclusion}

In this work we have discussed the main ingredients of the phonon dynamics
in the substrate that allows to understand the spatial profile of the SSE
signal. As we have shown, to explain the non-local effect, \textit{i.e.} its
dependence on the position of the probe along the substrate, one must
consider the spectral non-uniformity of the phonon distribution function
which can be {interpreted as spectrally} non-uniform temperature. A key
aspect of the non-locality is the explicit introduction of the boundaries
into the equations describing the propagation of {diffusing} phonons.

In addition, we have presented a scheme of the non-magnon mechanism in the
case when the ferromagnetic element of the device is conducting{\cite{ANE} and obtain the correct
magnitude of the effect. Furthermore, the spatial profile of the SSE signal, presented in Fig. \ref{fig_dt}, is very similar to the one shown as a 'universal' profile on the Fig. 2f of the Ref. \onlinecite{jaworski10}.}
Although the phonon kinetics at temperatures comparable with
$\theta_D$ is strongly modified by Umklapp processes, the measured
proportionality between the SSE signal and the magnitude of the
magnetization in Ref. \onlinecite{jaworski11} clearly indicates that near $%
T_{C}\approx 130$ K the effect is {still} dominated by the flux of the
spin-polarized electrons, instead of the magnon-mediated spin torque. We
believe that the difference between the data presented in Figs 2 and 3 of
Ref. \onlinecite{jaworski11} - in particular, the difference in the behavior
near the $T_{C},$ - supports this picture. Two different samples demonstrate
drastically different temperature behavior. The sample which is thicker and
grown on a substrate of a better quality has larger peak value of both $%
S_{xy}$ and thermopower $\alpha _{xx}$ and {also} much faster decay of $%
S_{xy}$ at approaching $T_{C}$. The stronger thermopower observed in the
thicker sample demonstrates that in this sample phonons lose momentum mainly
in collisions with electrons, while in the thinner sample, their scattering
on the defects is more efficient. However, due to strong sensitivity of the
phonon distribution function at the F-Pt boundary to the ratio $d_{F}/l_{in}^{F}\left( T\right)$, in the thicker sample the SSE decays with
temperature much faster than in the thinner one. As we have already
discussed, at large $d_{F}/l_{in}^{F}\left( T\right)$ the phonons
equilibrate before they reach the probe. Indeed, in the thicker sample (more
than three times thicker than the thinner one) the effect was not even
resolved near the Curie temperature within the accuracy of the measurement. This suggests the need to study the dependence of the SSE\ signal on the
thickness of the magnetic sample in otherwise identical conditions, \textit{i.e.}, keeping the properties of the insulating substrate and
semiconductor/substrate boundary the same.

\section*{Acknowledgements}

The authors are grateful for useful discussion with O. Tretiakov, J.
Heremans, R. Myers, and G. Jakob. We acknowledge the support by grants
ONR-000141110780, NSF-DMR-1006752, NSF-DMR-1105512, NHARP, and the Paul and
Tina Gardner fund for Weizmann-TAMU collaboration.

\section*{Author contributions}

{K. S. T. carried out theory calculations with assistance of
A. M. F and J. S. All three authors contributed equally to the project planning,
writing the text and analyzing the results.}

\section*{Competing financial interests}

The authors declare no competing financial interests.

\end{document}